\documentclass[aps,prl,twocolumn,superscriptaddress]{revtex4}
\usepackage{amsmath,amssymb}
\usepackage{graphicx}
\usepackage{bm}
\begin{document}

\preprint{LA-UR-08-04570}


\title{Microscopic study of the effect of impurities on the first order spin
density wave transition in BaFe$_2$As$_2$}


\author{S.-H. Baek}
\affiliation{Los Alamos National Laboratory, Los Alamos, NM 87545, USA}
\author{T. Klimczuk}
\affiliation{Los Alamos National Laboratory, Los Alamos, NM 87545, USA}
\affiliation{Faculty of Applied Physics and Mathematics, Gdansk University of
Technology, Narutowicza 11/12, 80-952 Gdansk, Poland}
\author{F. Ronning}
\affiliation{Los Alamos National Laboratory, Los Alamos, NM 87545, USA}
\author{E. D. Bauer}
\affiliation{Los Alamos National Laboratory, Los Alamos, NM 87545, USA}
\author{J. D. Thompson}
\affiliation{Los Alamos National Laboratory, Los Alamos, NM 87545, USA}
\author{N. J. Curro}
\affiliation{Department of Physics, University of California, Davis, CA 95616, USA}

\date{\today}

\begin{abstract}
We report an $^{75}$As NMR study of BaFe$_2$As$_2$ in both single crystals and
polycrystal forms.   We find that Sn impurities in the single crystal
dramatically alter the low energy spin fluctuations and suppress the ordering
temperature from 138 K to 85 K.  In contrast to Sn-free samples,  we find that
the temperature dependence of the $^{75}$As NMR spectra and spin lattice
relaxation rates reveal a second order phase transition to a state of
incommensurate magnetic order.
\end{abstract}

\pacs{76.60.-k, 75.30.Fv, 74.10.+v}

\maketitle

The newly discovered iron-based superconductors, RFeAsO$_{1-x}$F$_{x}$ (R=La,
Sm, Ce, Nd) and the doped ternary  compounds
$A_{1-x}$(K,Na)$_x$Fe$_2$As$_2$ ($A$ = Ba, Sr, Eu, and Ca) have
stimulated interest due to their similarities to the high $T_c$ cuprates and their
surprisingly high transition temperatures ($T_c=55$ K) \cite{kamihara08,
takahashi08,xchen08, ding08, ren08:2,ren08, gchen08,rotter08:2, ni08, wu08,
sasmal08, gchen08:2,ronning08}. Since these compounds do not contain the
ubiquitous CuO$_2$ planes present in the cuprate superconductors, they hold
promise to shed light on some of the mysteries of high-T$_c$ superconductivity
in transition metal compounds. The iron arsenides contain layers of FeAs
which are presumably responsible for metallic
behavior. The strongly correlated 3$d$-electrons of the Fe experience several
competing interactions that give rise to the simultaneous occurrence of a
spin-density wave (SDW) instability and a structural transition from tetragonal to
orthorhombic symmetry, as well as possible $d$-wave superconductivity
\cite{wu08, si08, grafe08}. The $A$Fe$_2$As$_2$ materials offer a unique
opportunity to probe the intrinsic physics at play in these materials, because
unlike the RFeAsO system, large single crystals of $A$Fe$_2$As$_2$ grow easily
and therefore are ideal for investigating the anisotropy and
2-dimensional (2D) electronic character that appear important for
superconductivity in the doped compounds.  In this Letter, we address the role
of impurities on the spin density wave transition in undoped BaFe$_2$As$_2$.

Investigations of the low energy spin dynamics of the RFeAsO materials reveal
slow magnetic fluctuations associated with a continuous magnetic phase
transition \cite{nakai08}.   We find similar magnetic fluctuations present in
BaFe$_2$As$_2$ (Ba122), which may play a crucial role in the pairing
mechanism for superconductivity \cite{monthoux07}.  However, the
details of these magnetic fluctuations appear to be highly sensitive to the
presence of lattice impurities.   Crystals of Ba122 grown in self-flux (FeAs)
have been reported to exhibit a sharp first-order like structural and magnetic
phase transition at 138 K to a commensurate magnetic structure
\cite{kitagawa08}. In contrast, we find that crystals grown in Sn flux exhibit a
transition at 85 K that is clearly second-order and incommensurate.  We argue
that random quenched Sn impurities at the level of 1\% in the structure
broaden the first-order transition, suppressing the ordering temperature and
completely rounding the phase transition \cite{imry79}.

In order to investigate the microscopic properties of the Ba122
system as well as to elucidate the difference between Sn-flux grown crystals
(SC) and nominally pure polycrystalline material (PC),
we performed $^{75}$As Nuclear Magnetic Resonance (NMR) on both SC and PC
samples of BaFe$_2$As$_2$.     Our measurements
provide unambiguous evidence for a magnetic transition at 85 K in the single
crystal. On the other hand, the
PC sample shows a transition at 138 K, which is
consistent with earlier reports \cite{rotter08:1, huang08}.  We find that the
low energy spin dynamics change drastically between samples, and the presence
of Sn impurities in the SC lead to an incommensurate magnetic structure.  The
extreme sensitivity of the spin dynamics to the presence of quenched random
impurities may play a crucial role in the development of superconductivity and
inhomogeneity in doped samples.

Polycrystalline samples of BaFe$_2$As$_2$ were prepared by first
synthesizing FeAs by reacting fine Fe powder and small pieces of
As in an alumina crucible in a sealed, evacuated silica ampoule.
The ampoule was slowly ($100 ^\circ$C/hr) heated to 600 $^\circ$C,
soaked for 4 hours, then heated to 700~$^\circ$C, soaked for 5 hours and
finally heated to 1050~$^\circ$C. That temperature was kept for 4
hours, and then the furnace was shut off. Next, a
stoichiometric mixture of FeAs and Ba were placed in a Ta tube,
which was sealed in an evacuated quartz ampoule. The temperature was
increased to 1050~$^\circ$C at 150~$^\circ$C/hr, held at temperature
for 36 hours, and then cooled to 800 $^\circ$C, where it was held
for 2 days.
Single crystals of BaFe$_2$As$_2$ were grown in Sn flux in the ratio Ba:Fe:As:Sn=1:2:2:20.
The starting elements were placed in an alumina crucible and sealed under
vacuum in a quartz tube.
The tube was placed in a furnace and heated to 500~$^{\circ}$C at 100~$^{\circ}$C,
held at that temperature for 6 hours.  This sequence was repeated at 750~$^{\circ}$C,
950~$^{\circ}$C and at a maximum temperature of 1100~$^{\circ}$C, with hold
times of 8 hr, 12 hr, and 4 hr,
respectively.  The sample was then cooled slowly ($\sim4 ^{\circ}$C/hr) to 600~$^{\circ}$C,
at which point the excess Sn flux was removed using a centrifuge.
The resulting plate-like crystals of typical dimensions $3\times3\times0.1$ mm$^3$
and are oriented with the $c$-axis normal to the plate.

\begin{figure}
\label{fig:1}
\centering
\includegraphics[width=0.45\textwidth]{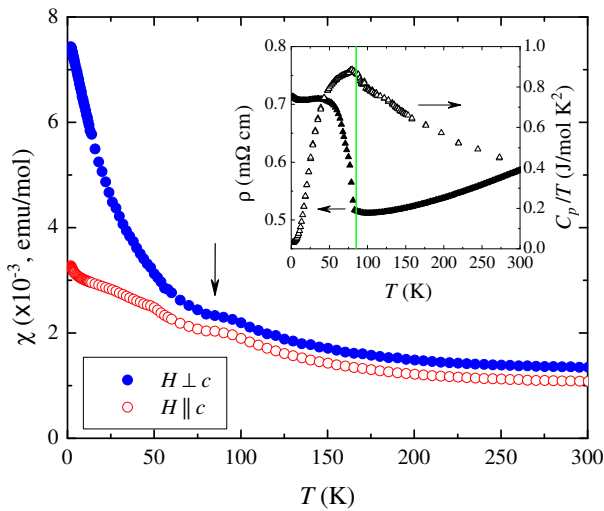}
\caption{Magnetic susceptibility $\chi(T)$ measured at $H=0.1$ T of SC BaFe$_2$As$_2$. A
small anomaly at 85 K (arrow) is clearly visible. Inset: Resistivity and
specific heat versus temperature in the SC sample.}
\end{figure}

Fig.~1 shows the magnetic susceptibility $\chi(T)$ of the SC sample measured at 0.1 T.
The temperature dependence and
anisotropy are similar to earlier reports, as are our measurements of resistivity and
specific heat \cite{ni08}. The rapid
increase of $\chi$ at low temperatures indicates the presence of a
small amount of an impurity phase,
which was not detectable in powder X-ray spectra. Relative to
observations in Ref.~\cite{ni08}, the impurity contribution to
$\chi$ is reduced by about 30~\% and, as a result, a small
anomaly at 85 K is clearly visible, suggesting a
magnetic phase transition. We point out that the $\chi(T)$ in the PC  \cite{rotter08:1}
and SC samples grown in FeAs flux \cite{wang08} actually
decreases slowly at high temperatures, and exhibit a clear drop at 140 K. The
relatively large impurity contribution at low temperatures in our case, however,
makes it difficult to obtain $\chi(T)$ of the pure Ba122 phase.
On the other hand, the As Knight shift, $K$,  probes microscopically the
intrinsic susceptibility of the Ba122. We find, in fact, that $K$ decreases
with decreasing temperature (Fig.~1) as observed in other iron arsenide materials.

\begin{figure}
\label{fig:2}
\centering
\includegraphics[width=0.45\textwidth]{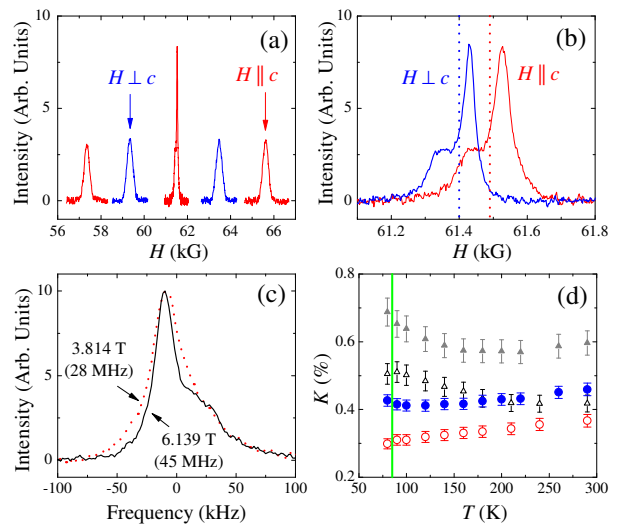}
\caption{ (a) Full $^{75}$As NMR spectra at a fixed frequency of 45
MHz and at 120 K. The satellite peaks (blue lines for $H \perp c$ and red
lines for $H \parallel c$) are equally spaced.
For simplicity, we omit the central line for $H \perp c$ which is 0.1 kG off
from the central line for $H \parallel c$.
(b) Central lines for both $H \perp c$ and $H \parallel c$. The vertical dotted lines
represent the midpoint between the satellites.
(c) Comparison of spectra in the SC sample for two different fields ($H \perp c$) at 220 K.
(d) The Knight shift of the primary (red, blue) site and the secondary site
(gray) versus temperature. The solid (open) gray triangles correspond to
$H\perp c$ ($H \parallel c$), and the vertical green line at 85 K corresponds
to $T_N$.}
\end{figure}

Fig.~2 (a) presents the $^{75}$As ($I=3/2$) NMR spectra of BaFe$_2$As$_2$  at 120
K, measured at constant frequency $f=45$ MHz. The spectrum shows the central
($I_z=+\frac{1}{2}\leftrightarrow -\frac{1}{2}$) transition, plus two satellite
peaks split by the quadrupolar interaction of the As nucleus with the local
electric field gradient (EFG).  We find that the quadrupolar frequency
$\nu_Q={3eQV_{zz}}/{2I(2I-1)h}=3(0.1)$
MHz, where $Q$ is the quadrupolar moment and $V_{zz}$ is the EFG.
Surprisingly, this value is 3--4 times smaller than
observed in the LaFeAsO compounds, but is consistent with other NMR reports in
BaFe$_2$As$_2$ \cite{grafe08, nakai08, mukuda08, kitagawa08}.  The spectra of
the central transition clearly reveal a narrow peak with approximately 60\% of
the total signal intensity, and a second broader peak consisting of $\sim 40$\%
of the intensity.  The two peaks arise from As nuclei in different local
charge (EFG) or magnetic environments.  In order to test whether the two peaks
arise from different EFGs, we compare two spectra at
different fields, since the second order shift to the central transition
frequency varies as $\nu_Q^2/H$ for $H\perp c$ \cite{slichter}.  As shown
in Fig.~2(c), the linewidth of the primary resonance
increases with decreasing $H$, whereas the secondary peak at higher
frequency is poorly resolved at the lower field, suggesting no significant
quadrupolar shift of this site.  This result suggests that the primary peak
experiences an EFG and is associated with the intrinsic Ba122 phase and the
secondary peak arises from As with little or no EFG.  Furthermore, we find that the two peaks have vastly different
magnetic shifts [Fig.~2(d)].   A plausible scenario, then, is that the secondary
peak arises from an impurity phase containing As, whereas the primary peak
arises from the BaFe$_2$As$_2$ phase, and probes the intrinsic physics of this
material.

\begin{figure}
\label{fig:3}
\centering
\includegraphics[width=0.45\textwidth]{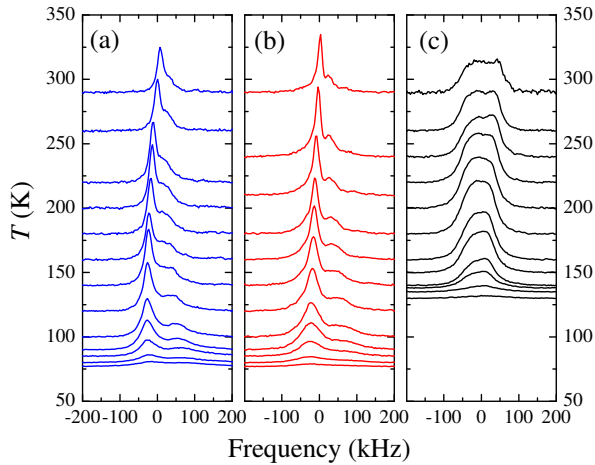}
\caption{$^{75}$As NMR spectra of the central transition in BaFe$_2$As$_2$ for both the SC and PC samples at 45 MHz for the case of : (a) $H \perp c$ at 6.139 T, (b) $H \parallel  c$ at 6.149 T, and (c) the polycrystal
sample at $H=6.148$ T. Two peaks are clearly distinguished in the SC. Below the ordering transition, the spectra
quickly broaden and the signal is drastically reduced.}
\end{figure}

The central line of $^{75}$As NMR ($\frac{1}{2} \leftrightarrow -\frac{1}{2}$)
and its temperature dependence are shown in Fig.~3(a) and (b).  Spectra measured on the
PC sample [Fig.~3(c)] reflect the Knight shift broadening expected for a
randomly oriented powder.
For the SC sample, we observe gradual broadening of spectra
down to 85 K and much faster broadening below 85 K. This is a
clear indication of a magnetic phase transition to an incommensurate state.
These data are consistent with a picture in which the internal field felt by
As nuclei is distributed spatially, giving rise to an
inhomogeneous broadening of the $^{75}$As NMR line.  This observation contrasts with recent
neutron scattering results and NMR in self-flux crystals of BaFe$_2$As$_2$ and
SrFe$_2$As$_2$ that
reveal commensurate order with $\mathbf{Q}=(1,0,1)$ \cite{huang08,jesche08,
kitagawa08}.  In the PC sample, we find that below 138 K the spectra also
broaden, showing no evidence for a commensurate magnetic order.  However, in
this case the spectra should be broadened anyway due to the strong angular
dependence of the resonance frequencies in the polycrystalline sample
\cite{fukazawa08}.

It is noteworthy that the spectra of the SC for both orientations do not show
any discontinuities through $T_N$.  This observation contrasts with NMR
measurements in self-flux grown BaFe$_2$As$_2$ and CaFe$_2$As$_2$ crystals
\cite{kitagawa08,baek08:1}.  In these cases, the EFG changes discontinuously at
the structural transition, and is reflected in the second order corrections to
the central transition frequency.  Our observations in the Sn-flux grown
crystal suggest there is no change in the structure down to 85 K.  Below this
temperature, the broad, incommensurate magnetic order preclude any detailed
analysis of the EFG through the NMR spectra.

\begin{figure}
\label{fig:4}
\centering
\includegraphics[width=0.45\textwidth]{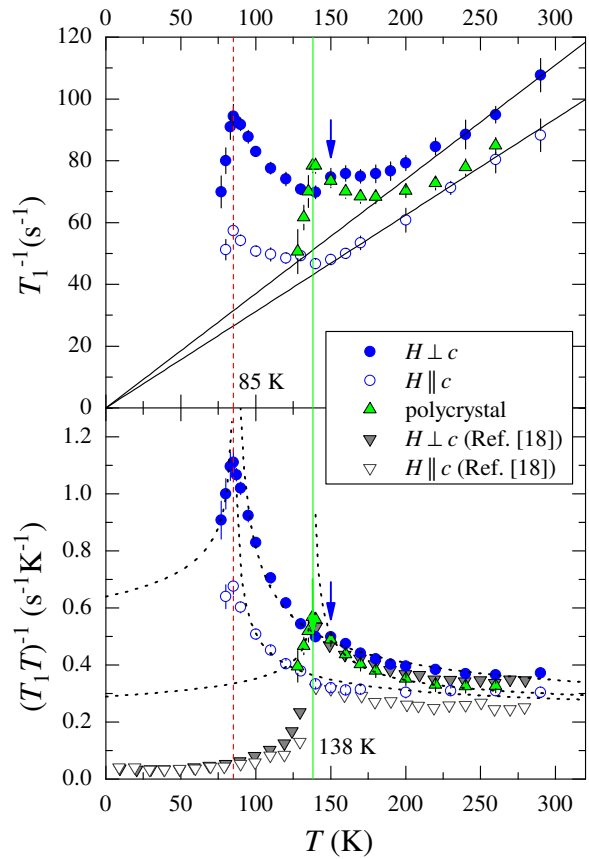}
\caption{(a) $T_1^{-1}$ versus $T$ in BaFe$_2$As$_2$ at 45 MHz for field along
$c$ and in the $ab$ plane.  The dotted red line at 85 K indicates the second
order transition (disordered sample) and the solid green line at 138 K
indicates the first order transition (pure sample). An additional weak maximum
at 150 K is evident only for  $H \perp c$. The solid lines are
guides to the eye.  (b) $(T_1T)^{-1}$ versus $T$. The dotted curves are fits
as described in the text, and the gray data points  are reproduced from
\cite{kitagawa08}.}
\end{figure}

In order to probe the low energy spin dynamics, we have  measured the spin
lattice relaxation rate, $T_1^{-1}$, as a function of temperature [Fig.~4(a)]
for both SC and PC samples. At high temperatures, $(T_1T)^{-1}$ [Fig.~4(b)]
approaches a constant
as expected for metallic systems. As the temperature decreases, $(T_1T)^{-1}$
increases to a maximum at $T_N=85$ K for the SC (135 K for the PC), and drops
rapidly below these temperatures.  This
behavior is similar to $^{139}$La NMR results on LaFeAsO at
$T_N=142$ K \cite{nakai08} and  $^{75}$As NMR on polycrystal
BaFe$_2$As$_2$ \cite{fukazawa08}, implying that the parent
compounds of superconducting materials are governed by similar
magnetic properties. The strong temperature dependence above $T_N$ reflects
the critical slowing down of spin fluctuations above a second order phase
transition, and the rapid drop below $T_N$ reflects the opening of the gap
associated with an SDW instability.
The dotted lines in Fig.~4 show fits to the self-consistent
renormalization (SCR) theory for weak itinerant antiferromagnets:
\begin{equation}
\frac{1}{T_1T} =
\begin{cases}
\label{eq:2}
    a +b/\sqrt{T-T_N}&   (T > T_N) \\
                  c\left(1-T/T_N\right)^{-\beta}&   (T < T_N),
\end{cases}
\end{equation}
where we have fixed the critical exponent for the order parameter, $\beta=0.15$,
as measured by Nakai et al.~\cite{moriya85, nakai08}. Here we find $a=0.18$
(sK)$^{-1}$,
$b=2.45$ s$^{-1}$K$^{-1/2}$ and $c=0.64$ (sK)$^{-1}$ for $H \perp c$; $a=0.22$
(sK)$^{-1}$, $b=1.0$ s$^{-1}$K$^{-1/2}$, and $c=0.29$ (sK)$^{-1}$ for $H ||c$; and  $a=0.2$ (sK)$^{-1}$, $b=1.2$ s$^{-1}$K$^{-1/2}$, and for the
PC sample.  These values are roughly one order of magnitude larger than those
reported for LaOFeAs
\cite{nakai08}. We note that the 2D character of the fluctuations is
clearly reflected in different $b$ values for the two orientations. It is
not clear if the difference arises from the anisotropy of the hyperfine
interaction itself, or rather in the intrinsic 2D fluctuations of the electronic system.

The striking feature in the spin lattice relaxation data is the
absence of any discontinuity associated with a first order transition.
Recent NMR measurements in a single crystal of BaFe$_2$As$_2$ grown by
FeAs flux show a drop in $T_1^{-1}$ at $T_N=135$ K \cite{kitagawa08}.  In
contrast, for crystals grown in Sn-flux, the structural and magnetic
transition is suppressed to 85 K.  We find that in this case, the phase
transition is no longer first order, but suggestive of a continuous second order
one.  The  $T_1^{-1}$ data for both samples agree at high temperature, but
clearly deviate below 135 K.  In the Sn-flux grown samples, $T_1^{-1}$
increases down to 85 K, representing the growth of slow antiferromagnetic
fluctuations associated with the \textit{second order} transition at 85 K.
Since Sn-flux grown samples are believed to contain Sn substitutions for As at
the level of $\sim$1 \%, we conclude that these Sn impurities dramatically
affect the first order transition \cite{ni08}. (Note that these impurities are present in the primary BaFe$_2$As$_2$ phase, and the effect is unrelated to the 40 \% extrinsic secondary phase discussed above.)  Not only do the impurities
suppress the ordering temperature by nearly 40\%, they also round out the
transition and alter the critical dynamics \cite{imry79, millis03}.
At a first order transition in the presence of disorder, local fluctuations of
the ordered phase may be destabilized because of the high cost of forming
phase boundaries.  Consequently the ordering temperature of the entire system
can be suppressed, and the discontinuity in the order parameter can be smeared
out. We suggest that the sensitivity of BaFe$_2$As$_2$ to Sn impurities
represents a generic aspect of the FeAs plane, and that similar effects may be
at play in the LaOFeAs materials.   This smearing out effect may explain the
strong increase in the order parameter and the  unusual $\beta=0.15$ exponent
observed in the ordered state of LaOFeAs \cite{nakai08,klauss08}.
Furthermore, recent $\mu$SR experiments suggest that the transition between an
SDW ground state and a superconducting one upon F doping is first order in
LaOFeAs \cite{luetkens08}, although there has been some report of coexisting SDW and
superconductivity for intermediate dopings \cite{drew08, hchen08}.

The dramatic temperature dependence of $T_1^{-1}$ above $T_N$ reflects the
critical slowing down of the spin fluctuations \cite{curro05}.  If
the magnetic transition is indeed smeared out by impurities, the critical
dynamics of this transition may not be universal, but rather are determined by
the extrinsic effect of the Sn impurities.  Nevertheless, these slow magnetic
dynamics may play an important role in the emergence of superconductivity
under pressure in the AFe$_2$As$_2$ compounds, or in F-doped LaOFeAs \cite{monthoux07}.

We thank
M. J. Graf, H.-J. Grafe, H.-H. Klauss and Q. Si for stimulating
discussions. Work at Los Alamos National Laboratory was performed under the
auspices of the US Department of Energy, Office of Science.

\bibliography{BaFe2As2_NMR}

\end{document}